# Resonant Raman scattering from polyacetylene and poly(*p*-phenylene vinylene) chains included into hydrogenated amorphous carbon

M. Rybachuk[1,3*], A. Hu[2], J.M. Bell[3]

[1] *Federal Institute for Materials Research and Testing (BAM), Division VI.4 Surface Technology, Unter den Eichen 87, 12205 Berlin, Germany*

[2] *Department of Physics, University of Waterloo, 200 Univ. Ave. West, Waterloo, ON, N2L 3G1, Canada*

[3] *Faculty of Built Environment and Engineering, Queensland University of Technology, 2 George St, Brisbane, Qld 4001, Australia*

ABSTRACT

The resonant Raman scattering in N-IR – UV range from amorphous hydrogenated carbon (*a*-C:H) reveal inclusions of trans-polyacetylene (*trans*-(CH)$_x$) chains with approximate length of up to 120 *C=C* units and inclusions of poly(*p*-phenylene vinylene) (PPV) polymer chains. The PPV is evidenced by a strong dispersive mode at *ca.* 1175 cm$^{-1}$. It was found that the Raman response from core $A_g$ *trans*-(CH)$_x$ modes incorporated into *a*-C:H to changing excitation energy is identical to of free-standing chains thus facilitating identification of *trans*-(CH)$_x$ in complex carbonaceous materials spectra.

MAIN TEXT

It is known that diamond-like carbon (DLC) can host a basic polymer, the *trans* isomer of polyacetylene (*trans*-(CH)$_x$) initially reported for CVD grown diamond[1] and later found in low temperature grown hydrogenated amorphous carbon (*a*-C:H) films[2]. Excellent conductivity of *trans*-(CH)$_x$ due to strong electron-phonon (*e*-ph) and electron-electron coupling originating from delocalised $\pi$ electrons and an effective lattice nonlinearity[3,4] and the large third-order nonlinear optical susceptibility that allows the chain to withstand high peak pump powers without damage to the sample, ensure considerable interest in this polymer as a non-linear optical material[5]. Achieving controlled inclusion of *trans*-(CH)$_x$ into host DLC has been difficult and only short ($\leq 20$ of *C=C* units) *trans*-(CH)$_x$ segments have been found to date[1,6]. Recently, Hu *et al.*[7,8] demonstrated that variably bonded carbon atoms, including *trans*-(CH)$_x$, can be incorporated on a carbon surface using ultra-short laser pulses. Apart from *trans*-(CH)$_x$ segments DLC can also contain nanoparticles like carbon onions[9] or spherical nanocrystallites as reported by Chen *et al.*[10]. These greatly reduce internal stress and thus are favourable for tribological applications.

We present here a resonant Raman scattering (RRS) investigation of *a*-C:H films synthesised in a low temperature inductively coupled plasma (ICP) reactor[11]. Although films are indeed of low stress and host *trans*-(CH)$_x$ chains of significant length ($\leq 120$ of *C=C* units), they also contain poly(*p*-phenylene vinylene) (PPV) inclusions that have not been reported previously. The RRS technique probes atomic configurations in materials via the vibrational density of states[3,6,12] and in this work laser excitation energies, $\hbar\omega_L$ ranging from 1.58 eV (N-IR) to 5.08 eV (UV) are used, ensuring bonding and structural disorder in the great majority of $sp^3$, $sp^2$ and $sp$ carbon mixtures are studied. We also demonstrate that the response of *trans*-(CH)$_x$ segments in *a*-C:H to changing excitation energy is identical to that of free-standing isolated *trans*-(CH)$_x$ chains, both empirically and theoretically, using either the bi-modal distribution model proposed by Brivio *et al.*[13] or the amplitude mode theory proposed by Ehrenfreund *et al.*[3]. Our findings exemplify an approach which facilitates the extraction of *trans*-(CH)$_x$ contributions from the core *a*-C:H, DLC or carbonaceous materials spectra thus precluding overfitting as in case of Piazza *et al.*[2].

*a*-C:H films were deposited on *Si* at the rate of ~30 nm/hour using *CH$_4$/Ar* plasma in Helmholtz type ICP reactor[11] at temperatures of ≤ 400 K as described elsewhere[14]. The deposition pressure was ~6×10$^{-2}$ Pa and the substrate was negatively DC biased at 250-300 V. The fabricated films were of low stress ≤ 1 GPa, with hardness of ≤ 20 GPa and a friction coefficient of 0.07 at 70 % humidity as measured by nano-mechanical testing (UMIS). Electrical resistivity was ≥ 8×10$^8$ Ω cm. Films were ~140 nm thick with a maximum refractive index of 2.2 in the UV-blue region measured by IR-UV spectroscopic ellipsometry (J.A. Woollam Co.) The hydrogen content was found to be 27.5 (± 2.5) at. % for all films as determined from Fourier Transform infrared (FT-IR) spectroscopy (Nicolet Nexus). Analysis of $C_{1s}$ and valence bands of X-ray photoelectron spectra (Kratos Axis Ultra) determined the *sp*, *sp$^2$* and *sp$^3$* contents to be 2, 68 and 30 % respectively with the uncertainty of 1.25 %. The *sp*-hybridised content was verified using Raman and FT-IR, and the *sp$^3$* content using 244 nm Raman results[12]. Unpolarised Raman spectra (5.08 - 1.58 eV) were obtained *ex situ* at 293 K using 244, 532, 633 and 785 nm Renishaw instruments and 325 nm and 442 nm Kimmon Raman instruments. All excitation wavelengths excluding 785 nm were pulsed; the 785 nm was a continuous wavelength laser source. The frequency-doubled *Ar* ion laser was used for 244 nm, *He/Cd* for 325 nm and 442 nm, the frequency-doubled YAG laser was used for 532 nm, *He/Ne* gas laser was used for 633 nm and a diode laser source was used for 785 nm excitations. All measurements were taken in dynamic mode where a specimen is moved linearly at speeds of ≤ 30 µm/s and laser power was kept < 1 mW minimizing thermal damage.

Fig. 1 shows RRS spectra of an *a*-C:H film. After a linear background subtraction the spectra were all fitted with Gaussian line-shapes using a nonlinear least squares fitting[15]. Fitted bands are the common DLC *D* and *G* modes (N-IR and visible) and *T* mode (UV)[12] and the two $A_g$ zone center vibrational modes of *trans*-$(CH)_x$[3,4,13]: the *C-C* $\omega_1$ at ~1060 cm$^{-1}$, and the *C=C* backbone stretching $\omega_3$ mode at ~1450 cm$^{-1}$. The weak $\omega_2$ mode at ~1280 cm$^{-1}$ was not detectable though its contributions may be obscured by the *D* and $\omega_1$ bands. The absorption for bulk *trans*-$(CH)_x$ occurs at 1.5 - 1.7 eV[4,13] and corresponds to positions of the $A_g$ zone centres at 1060, 1280 and 1450 cm$^{-1}$. This applies for N-IR excitation. As $\hbar\omega_L$ increases moving away from resonance, shoulders appear at the high frequency side of the $\omega_1$ and $\omega_3$ modes, eventually developing into secondary peaks[3,13,16] at excitation energies well above the band gap of 2.71 eV[4]. The RRS spectra disperse[6] and these peaks change in intensity (*I*) and widths (*Γ*). The complexity of separating intercalated *trans*-$(CH)_x$ from the host DLC modes lead us to analyse a single symmetric band distribution. This proved to be sufficient[3] to account for a double peak Raman structure.

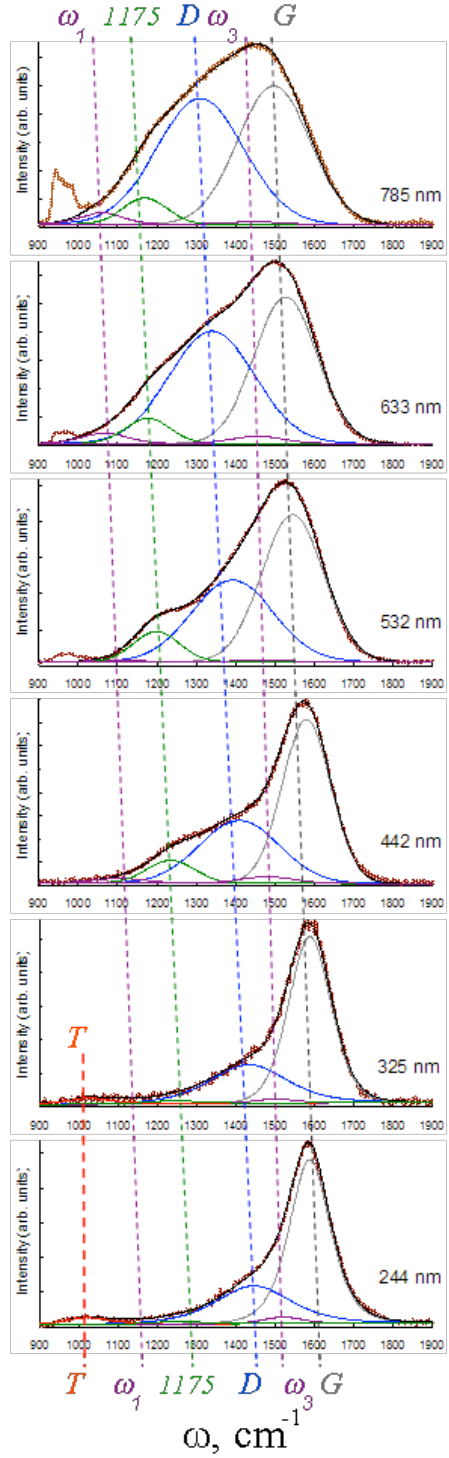

FIG. 1. The RRS spectra of examined *a*-C:H films showing contributions from *trans*-(CH)$_x$ ($\omega_1$ and $\omega_3$ modes), PPV (1175 cm$^{-1}$ mode) and DLC (*D*, *G* and *T* modes). An asymmetric peak visible at N-IR – visible (green) $\hbar\omega_L$ at ~950 cm$^{-1}$ is the second order *Si*.

A peak positioned at 1175 cm$^{-1}$ at N-IR $\hbar\omega_L$ we assign to a *CC–H* bending mode of the ring in neutral poly(*p*-phenylene vinylene)[17-19]. The origin of this mode could be due to introduction of heteroatoms (defects) in *sp²* rings since in single crystals these lead to a relaxation of wave vector *k*=0 selection rule[6,12] thus providing a mechanism for phonons from outside the centre of the Brillouin zone to contribute to the Raman scattering. Introduction of heteroatoms allows delocalisation of π electrons confined to the *sp²* rings and thus dispersion[12,17].

Other PPV zone centre vibrational modes should be positioned at higher frequencies in the ranges[18,19] 1200 – 1330 and 1540 – 1625 cm$^{-1}$, but these are certainly obscured by the host *D* and the *G* modes. The large width of the 1175 cm$^{-1}$ mode suggests a combination of a vinylene and a *CC–H* ring bend modes since the zone mode frequency for vinylene[20] is at 1145 cm$^{-1}$.

As $\hbar\omega_L$ energy increases all peaks shift to a higher frequency; DLC modes are obeying phonon confinement rules[12], Fig. 2 (a) shows peak dispersion, $\Delta\omega$, the shift in peak position relative to the N-IR excitation peak position. Fig. 2 (b) summarizes changes in $\Gamma$ for all fitted peaks. The steady *I(D)/I(G)* ratio decrease from ~0.9 to 0.2, pronounced reduction in $\Gamma_D$ and $\Gamma_G$ and the *G* peak saturation[12] at ~1590 for 244 nm excitation are indicative of a highly ordered and symmetric *sp²* phase[12,14]. The band gap for PPV is 2.2 – 2.3 eV[18] and that is selectively probed by a resonance frequency of green 532 nm laser; Fig. 1 shows the elevated intensity, $I_{1175}$ and Fig. 2 (b) the broadening peak width, $\Gamma_{1175}$ for the PPV peak. This peak is almost certainly of *sp²* origin since its contributions disappear in UV excitation. There is an increase in *I(ω₃)/I(ω₁)* intensity ratio (Fig. 1) and in peak widths (Fig. 2 (b)) for *trans*-(CH)$_x$ $\omega_1$ and $\omega_3$ peaks that become transformed when the $\hbar\omega_L$ exceeds the band gap (~1.5 eV[4,13]) indicative of resonant probing of an inhomogeneous chain. Our results show that inhomogeneity of intercalated *trans*-$(CH)_x$ chains measured using the distribution of the *e*-ph coupling constant $\lambda$, $p(\lambda)$ of the amplitude mode (AM) theory proposed by Ehrenfreund *et al.*[3] gives $\lambda \sim 0.17$ for N-IR and $\sim 0.24$ for UV; in good agreement with the AM model. $\lambda$ determines the Peierls relation for the energy gap and its distribution arises from finite localisation lengths and bond length disorder. The AM results indicate that *trans*-$(CH)_x$ chains probed by high $\hbar\omega_L$ are of shorter $\pi$-conjugation lengths and of higher bond disorder. The approximate chain lengths for both single *C-C* and double *C=C* bonds of *trans*-$(CH)_x$ segments were determined using the bi-modal distribution model proposed by Brivio *et al.*[13] and was found to be ~120 of bond lengths units (N-IR), at the estimation limit of the model, and with a population of short chain of approximately 8 (UV). Shorter chains are probed by higher $\hbar\omega_L$. The average chain population is ~25 (± 5) bond length units owing to the uncertainties given by the Raman fitting and the bi-modal distribution model[13]. All *trans*-$(CH)_x$ chains are highly disordered as evidenced by wide $\omega_1$ and $\omega_3$ Raman peaks reaching their maximum in the blue-green range.

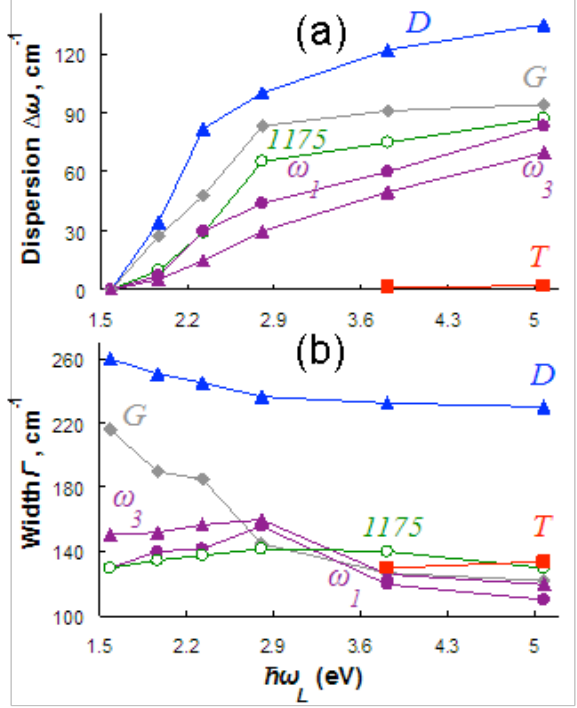

FIG. 2. (a) Peak dispersion, $\Delta\omega$ and (b) peak widths, $\Gamma$ for all constituent peaks as a function of the laser excitation energy $\hbar\omega_L$.

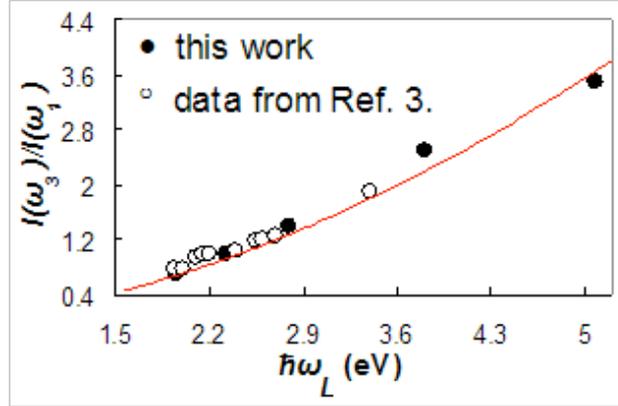

FIG. 3. The intensity ratio of $I(\omega_3)/I(\omega_1)$ vs. the laser excitation energy $\hbar\omega_L$ for *trans*-$(CH)_x$ inclusions in *a*-C:H. Solid line is a theoretical calculation performed using the amplitude mode formalism[3].

We have extended the $I(\omega_3)/I(\omega_1)$ vs. $\hbar\omega_L$ theoretical AM distribution calculations (independent of chain length) of Ehrenfreund *et al.*[3] for the visible range to include N-IR and UV $\hbar\omega_L$. Fig. 3 shows that our experimental results are in good agreement with the theoretical prediction and with Ehrenfreund's experimental data; clearly both the free-standing and incorporated *trans*-$(CH)_x$ chains obey the same $I(\omega_3)/I(\omega_1)$ evolution formalism.

Long *trans*-$(CH)_x$ chains and PPV inclusions are only possible in an ordered $sp^2$ *a*-C:H matrix that is achieved via deposition in ICP reactor analogous to used by Chen *et al.*[10] with high plasma density and low electron temperature compared to conventional DLC deposition systems.

In summary, the RRS investigation of ICP fabricated *a*-C:H films showed that films host long *trans*-$(CH)_x$ chains with up to 120 *C=C* bond length units and also poly(*p*-phenylene vinylene) as evidenced by the 1175 cm$^{-1}$ Raman mode. We have postulated the origin of this PPV mode and provided a theoretical basis for arguing the response of *trans*-$(CH)_x$ chains in the *a*-C:H matrix to changing Raman excitation energy is identical to of free-standing chains. The evolution of relative intensity ratio for core *trans*-$(CH)_x$ modes will facilitate identification of *trans*-$(CH)_x$ modes in other complex carbonaceous materials spectra.

ACKNOLEGEMENTS

This work was supported by the Australian Research Council (LP0235814) and BAM research fellowship funding.

REFERENCES

¹ T. López-Ríos, É. Sandré, S. Leclercq, and É. Sauvain, Phys. Rev. Lett. **76,** 4935 LP (1996).

² F. Piazza, A. Golanski, S. Schulze, and G. Relihan, Appl. Phys. Lett. **82,** 358 (2003).

³ E. Ehrenfreund, Z. Vardeny, O. Brafman, and B. Horovitz, Phys. Rev. B **36,** 1535 LP (1987).

⁴ A. J. Heeger, S. Kivelson, J. R. Schrieffer, and W.-P. Su, Rev. Modern Phys. **60,** 781 LP (1988).

⁵ A. J. Heeger, Rev. Modern Phys. **73,** 681 (2001).

⁶ A. Ferrari and J. Robertson, Phys. Rev. B **63,** 121405 (2001).

⁷ A. Hu, M. Rybachuk, Q. B. Lu, and W. W. Duley, Appl. Phys. Lett. **91,** 131906 (2007).

⁸ A. Hu, Q.-B. Lu, W. W. Duley, and M. Rybachuk, J. Chem. Phys. **126,** 154705 (2007).

⁹ G. A. J. Amaratunga, M. Chhowalla, C. J. Kiely, I. Alexandrou, R Aharonov, and R. M. Devenish, Nature **383,** 321 (1996).

¹⁰ L.-Y. Chen and F. C.-N. Hong, Appl. Phys. Lett. **82,** 3526 (2003).

¹¹ I. K. Varga, J. Vacuum Sci. Tech. A **7,** 2639 (1989).

¹² A. C. Ferrari and J. Robertson, Phys. Rev. B **64,** 075414 (2001).

¹³ G. P. Brivio and E. Mulazzi, Phys. Rev. B **30,** 876 (1984).

¹⁴ M. Rybachuk and J. M. Bell, Diamond Rel. Mat. **15,** 977 (2006).

¹⁵ D. C. Benner, C. P. Rinsland, V. M. Devi, M. A. H. Smith, and D. Atkins, J. Quant. Spec. Radiat. Transfer **53,** 705 (1995).

16   D. B. Fitchen, Molecul. Cryst. Liq. Cryst. **83,** 95 (1982).

17   V. Hernandez, C. Castiglioni, M. Del Zoppo, and G. Zerbi, Phys. Rev. B **50,** 9815 (1994).

18   M. Tzolov, V. P. Koch, W. Bruetting, and M. Schwoerer, Synth. Metals **109,** 85 (2000).

19   I. Orion, J.-P. Buisson, and S. Lefrant, Phys. Rev. B **57,** 7050 (1990).

20   M. Baitoul, J. Wery, J.-P. Buisson, G. Arbuckle, H. Shah, S. Lefrant, and M. Hamdoume, Polymer **41,** 6955 (2000).